# Ultrathin epitaxial ferroelectric films grown on compressive substrates: Competition between the surface and strain effects

A. G. Zembilgotov

*(State Technical University of St. Petersburg, St. Petersburg, Russia)*

N. A. Pertsev, H. Kohlstedt, and R. Waser

*(Institut für Festkörperforschung, Forschungszentrum Jülich, Jülich, Germany)*

The mean-field Landau-type theory is used to analyze the polarization properties of epitaxial ferroelectric thin films grown on dissimilar cubic substrates, which induce biaxial compressive stress in the film plane. The intrinsic effect of the film surfaces on the spontaneous polarization is taken into account via the concept of the extrapolation length. The theory simultaneously allows for the influence of the misfit strain imposed on the film lattice by a thick substrate. Numerical calculations are performed for $PbTiO_3$ and $BaTiO_3$ films under an assumption of the polarization reduction in surface layers. The film mean polarization is calculated as a function of film thickness, temperature, and misfit strain. It is shown that the negative intrinsic size effect is reduced in epitaxial films due to the in-plane compression of the film lattice. At room temperature, strong reduction of the mean polarization may take place only in ultrathin films (thickness ~ 1 nm). Theoretical predictions are compared with the available experimental data on polarization properties of $BaTiO_3$ films grown on $SrRuO_3$ coated $SrTiO_3$.

## I. INTRODUCTION

In view of the sustained trend towards further miniaturization of microelectronic devices, investigations of the scaling effects in electronic materials like ferroelectrics acquire great practical importance. Among these, the size effect on ferroelectricity in thin epitaxial layers is of special interest because of the use of ferroelectric films in modern nonvolatile random access memories (FeRAMs).[1,2] To ensure write and read operations of the FeRAM device, ferroelectric layers should show large switchable polarizations even at a small thickness, which is necessary to minimize the volume occupied by a ferroelectric cell. Therefore, it is important to study the dependence of the ferroelectric polarization on the film thickness both experimentally and theoretically.

The intrinsic surface effect on phase transitions in ferroelectric thin films and their physical properties is a subject of theoretical investigations since early 80-ties.[3-10] This effect was regarded as a possible cause of the suppression of ferroelectricity in ultrathin films and its disappearance at a nonzero critical thickness. The magnitude of this critical thickness has been calculated in the framework of the thermodynamic theory based on the concept of extrapolation length. For $BaTiO_3$ and $PbTiO_3$ films, for example, the critical thickness at room temperature was estimated to be about 20 nm and 5.5 nm, respectively.[8] However, the theoretical calculations were performed for free standing films, but not for epitaxial films grown on much thicker substrates, which are usually studied experimentally. At the same time, it was shown recently[11] that the two-dimensional clamping and straining of a ferroelectric layer by a dissimilar thick substrate may affect considerably its phase state and physical properties.

Besides, some recent experiments demonstrated that even ultrathin epitaxial films of perovskite ferroelectrics may possess pronounced ferroelectric properties. For instance, it was found that 4-nm-thick $Pb(Zr_{0.2}Ti_{0.8})O_3$ films may exhibit a stable polarization orthogonal to the surface.[12] Moreover, 12-nm-thick $BaTiO_3$ films grown on $SrTiO_3$ displayed a remanent polarization larger than the bulk spontaneous polarization,[13] which is in qualitative agreement with the predictions of the thermodynamic theory of "thick" epitaxial films.[11]

The above facts motivated us to develop a phenomenological theory, which takes into account both the two dimensional clamping and straining of epitaxial ferroelectric layers and the intrinsic surface effect on their polarization properties. In this paper, we analyze ferroelectric films with the polarization orthogonal to the substrate and a positive extrapolation length. This situation corresponds to the films of perovskite ferroelectrics grown on "compressive" substrates (i.e. inducing compressive in-plane stresses in the film) and characterized by a polarization reduction in subsurface layers.

In Section II the basic thermodynamic relations are formulated. Variations of the mean polarization with the film thickness and temperature are discussed in Section III. The strain effect on the transition temperature and polarization properties of ultrathin films is described in Section IV, where the comparison of theoretical results with available experimental data is also given.

## II. THERMODYNAMICS OF THE OUT-OF-PLANE POLARIZATION STATE IN ULTRATHIN FILMS

We shall consider epitaxial thin films of perovskite ferroelectrics, which are grown in a high-temperature paraelectric state on a thick (001)-oriented cubic substrate. In the ferroelectric phase, which forms in a film during the cooling from the deposition temperature $T_g$, the polarization orientation generally depends on the misfit strain $S_m$ in the epitaxy.[11,14] [$S_m = (b^*-a_0)/b^*$, where $b^*$ is the substrate effective lattice parameter,[15] and $a_0$ is the equivalent cubic cell constant of the free standing film.] In a wide range of misfit strains and temperatures, however, the polarization state with the spontaneous polarization **P**$_s$ *orthogonal* to the substrate surface is expected to be stable. The stability conditions of this out-of-plane polarization state (tetragonal *c* phase) depend on the electrostrictive constants $Q_{ij}$ of the paraelectric phase. In ferroelectric substances like BaTiO$_3$ (BT) and PbTiO$_3$ (PT), which have $Q_{11} > 0$ and $Q_{12} < 0$, the *c* phase is stable at *negative* misfit strains satisfying the inequality $S_m < S_m^0(T) = Q_{12} P_0^2(T)$, where $P_0$ is the spontaneous polarization of a free crystal.[14]

For the correct thermodynamic description of epitaxial thin films, an appropriate form of the free-energy function must be chosen, which corresponds to actual mechanical and electric boundary conditions of the problem. In this paper, we shall consider the usual situation, where the film/substrate system is not subjected to *external* mechanical forces. We also assume that the film is sandwiched between extended identical electrodes in the short-circuited condition. Since there is no work done by extraneous mechanical sources, the total free energy $\Im$ of the epitaxial layer can be derived solely from the spatial distribution of the Helmholtz free-energy density $F$ within the film.

For perovskite crystals that are proper ferroelectrics and improper ferroelastics, the energy density $F$ may be approximated by a polynomial in the polarization components $P_i$, their first gradients $P_{i,j}$, and lattice strains $S_n$.[16] In the particular case of the *c* phase with the tetragonal polar $x_3$ axis orthogonal to the substrate ($P_1 = P_2 = 0$, $P_3 \neq 0$), this polynomial reduces to

$$F = \alpha_1 P_3^2 + \alpha_{11} P_3^4 + \alpha_{111} P_3^6 + \frac{1}{2} c_{11}(S_1^2 + S_2^2 + S_3^2) + c_{12}(S_1 S_2 + S_1 S_3 + S_2 S_3) - q_{11} S_3 P_3^2 - q_{12}(S_1 + S_2) P_3^2 + \frac{1}{2} g_{11} P_{3,3}^2, \quad (1)$$

where $\alpha_1$, $\alpha_{11}$, and $\alpha_{111}$ are the dielectric stiffness and higher-order stiffness coefficients at constant strain, $q_{ij}$ are the relevant electrostrictive constants, and $c_{ij}$ are the film elastic stiffnesses at constant polarization. Equation (1) assumes that the polarization changes only in the film thickness direction, i.e. along the $x_3$ axis, and ignores the depolarizing field, which might be associated with the gradient $P_{3,3}$. (This field should be negligible in perovskite ferroelectrics due to their finite conductivity.[17]) It also allows for the absence of shear strains $S_4$, $S_5$, and $S_6$ in a tetragonal film grown on a cubic substrate.[18] It should be noted that Eq. (1) neglects the contribution of strain gradients, which may be important in proper ferroelastics,[19,20] and the coupling between these gradients and polarization.

Using the mechanical boundary conditions of the problem, we can eliminate lattice strains from the free-energy expansion (1). Since the substrate is usually several orders of magnitude thicker than the epitaxial layer, the in-plane strains $S_1$ and $S_2$ must be constant throughout the film thickness and equal to the misfit strain $S_m$ in the heterostructure. On the other hand, the normal stress $\sigma_3$ is zero on the film free surface and satisfies the equation of mechanical equilibrium $\sigma_{3,3} = 0$. Therefore, the condition $\sigma_3 = 0$ holds throughout the film thickness, which enables us to calculate the strain $S_3$ via the thermodynamic relation $\sigma_3 = \partial F / \partial S_3$ as $S_3(x_3) = [q_{11} P_3^2(x_3) - 2 c_{12} S_m]/c_{11}$. Substituting $S_n$ into Eq. (1), we obtain

$$F = \frac{c_{11}^2 + c_{11} c_{12} - 2 c_{12}^2}{c_{11}} S_m^2 + \left(\alpha_1 + 2 \frac{q_{11} c_{12} - q_{12} c_{11}}{c_{11}} S_m\right) P_3^2 + \left(\alpha_{11} - \frac{q_{11}^2}{2 c_{11}}\right) P_3^4 + \alpha_{111} P_3^6 + \frac{1}{2} g_{11} P_{3,3}^2. \quad (2)$$

Since for perovskite ferroelectrics the Gibbs energy function $G$ is defined better than the function $F$,[21,22] it is convenient to express parameters of Eq. (2) in terms of the material constants involved in the expansion of $G$. This can be done by deriving the Helmholtz energy function $F$ via the inverse Legendre transformation of $G$ and comparing it the with the standard form of $F$. The substitution of the derived relationships into Eq. (2) after some mathematical manipulation yields (in the contracted notation $P_3 \equiv P$, $x_3 \equiv z$)

$$F(z) = \frac{S_m^2}{s_{11} + s_{12}} + a_3^*(S_m) P^2 + a_{33}^* P^4 + a_{111} P^6 + \frac{1}{2} g_{11} \left(\frac{dP}{dz}\right)^2, \quad (3)$$

$$a_3^* = a_1 - S_m \frac{2 Q_{12}}{s_{11} + s_{12}}, \quad (4)$$

$$a_{33}^* = a_{11} + \frac{Q_{12}^2}{s_{11} + s_{12}}, \quad (5)$$

where $a_1$, $a_{11}$, and $a_{111}$ are the dielectric stiffness and higher-order stiffness coefficients at constant stress,[22] $Q_{ij}$ are the electrostrictive constants related to $G$, and $s_{ij}$ are the elastic compliances at constant polarization. It should be emphasized that the renormalized coefficients $a_3^*$ and $a_{33}^*$ given by Eqs. (4)-(5) coincide with the corresponding coefficients of the modified thermodynamic potential $\widetilde{G}$, which was introduced in Ref. 11 for the thermodynamic description of homogeneously polarized epitaxial films.

The film free energy $\Im/A$ per unit area can be calculated now by integrating the density (3) over the film thickness $H$ and adding the contributions associated with the film surfaces.[3] This yields

$$\Im/A = \int_{-H/2}^{H/2} F(z) dz + \frac{1}{2} g_{11} \delta^{-1} \left(P_-^2 + P_+^2\right), \quad (6)$$

where $\delta$ is the so-called extrapolation length, and $P_+$ and $P_-$ are the polarization values on the film surfaces, i.e. at $z = +H/2$ and $z = -H/2$, respectively. To find the equilibrium polarization profile $P(z)$, we may use the Euler-Lagrange equation resulting from Eq. (6), which has the form

$$g_{11}\frac{d^2P}{dz^2} = 2a_3^*P + 4a_{33}^*P^3 + 6a_{111}P^5 \qquad (7)$$

The boundary conditions corresponding to the surface terms in Eq. (6) read[3]

$$\frac{dP}{dz} = \pm\delta^{-1}P(z) \quad \text{at} \quad z = \mp H/2. \qquad (8)$$

Since only films with identical electrodes are considered here and the mechanical substrate effect does not produce asymmetry of the polarization profile $P(z)$, we may assume the latter to be symmetric with respect to the film center $z = 0$. This yields $P_+ = P_- = P_b$, $P(z = 0) = P_c$, and $dP/dz = 0$ at $z = 0$, so that the first integration of Eq. (7) gives

$$\frac{1}{2}g_{11}\left(\frac{dP}{dz}\right)^2 = a_3^*(P^2 - P_c^2) + a_{33}^*(P^4 - P_c^4) + a_{111}(P^6 - P_c^6) . \qquad (9)$$

Using the boundary condition (8) together with Eq. (9), we find the relation between $P_c$ and $P_b$ as

$$\frac{g_{11}}{2\delta^2}P_b^2 = a_3^*(P_b^2 - P_c^2) + a_{33}^*(P_b^4 - P_c^4) + a_{111}(P_b^6 - P_c^6) . \qquad (10)$$

The second integration of Eq. (7) at $0 \le z \le H/2$ yields

$$z = -\int_{P_c}^{P(z)} \frac{\sqrt{g_{11}/2}\, dp}{\sqrt{a_3^*(p^2 - P_c^2) + a_{33}^*(p^4 - P_c^4) + a_{111}(p^6 - P_c^6)}}, \qquad (11)$$

where the extrapolation length $\delta$ is taken to be *positive* in order to describe the worst case of the polarization suppression in the surface layers.[4-6] By solving Eqs. (10) and (11) simultaneously, we can calculate the polarization $P(z)$ in a ferroelectric slab situated at a given distance $z$ from the film center. This enables us to determine the polarization profile $P(z)$ within the film. Finally, the mean polarization $\overline{P}$ of the film can be found as

$$\overline{P} = \frac{2}{H}\int_0^{H/2} P(z)dz . \qquad (12)$$

### III. FILM POLARIZATION: THICKNESS AND TEMPERATURE DEPENDENCES

Using the thermodynamic approach described in Sec. II, it is possible to calculate the mean polarization $\overline{P}$ in an epitaxial layer as a function of the film thickness $H$, temperature $T$, and the misfit strain $S_m$ in the heterostructure. In general, Eqs. (10)-(11) can be solved only numerically for a set of material parameters corresponding to a concrete ferroelectric substance. However, some universal relationships can be obtained in the so-called $P^4$-approximation, which neglects the sixth-order polarization term in the free energy expansion (3). These relationships correspond to epitaxial films experiencing a *second-order* paraelectric to ferroelectric phase transition during the cooling from the deposition temperature. It should be noted that, due to the two-dimensional clamping of the film on a thick substrate, the ferroelectric transformation in an epitaxial layer may become the second-order phase transition, even if it is of the first order in a free bulk crystal.[11] This situation takes place in epitaxial single-domain BT and PT films,[11] and we shall check the validity of the $P^4$-approximation for these films below.

Neglecting the sixth-order term in Eq. (3), we can write the Euler-Lagrange equation (7) and the boundary condition (8) in the following normalized form:

$$\frac{d^2\rho}{d\zeta^2} = -2\rho + 2\rho^3, \qquad (13)$$

$$\frac{d\rho}{d\zeta} = \pm\frac{\xi^*}{\delta}\rho_b \quad \text{at} \quad \zeta = \mp\frac{H}{2\xi^*}, \qquad (14)$$

where $\rho = P/P_s$, $\rho_b = P_b/P_s$, and $\zeta = z/\xi^*$. Here we introduced a modified ferroelectric correlation length $\xi^*$, which is given by the relation

$$\xi^* = \sqrt{\frac{g_{11}}{|a_3^*|}}, \qquad (15)$$

and differs from the correlation length $\xi = \sqrt{g_{11}/|a_1|}$ of a mechanically free bulk crystal (due to the strain effect, see Sec. IV). The polarization $P_s$ used in the normalization procedure corresponds to the spontaneous polarization of a thick epitaxial layer ($H \gg \xi^*$), where the surface effect can be neglected so that

$$P_s^2 = -\frac{a_3^*}{2a_{33}^*} . \qquad (16)$$

For the symmetric polarization profile, the integration of Eq. (13) yields ($\rho_c = P_c/P_s$)

$$\zeta = -\int_{\rho_c}^{\rho(\zeta)} \frac{dp}{\sqrt{p^4 - \rho_c^4 - 2(p^2 - \rho_c^2)}}, \qquad (17)$$

$$\rho_b^2 = \left(1 + \frac{\xi^{*2}}{2\delta^2}\right) - \sqrt{\left(1 + \frac{\xi^{*2}}{2\delta^2}\right)^2 + \rho_c^4 - 2\rho_c^2} . \qquad (18)$$

Equations (17)-(18) implicitly define the function $\rho(\zeta)$ and demonstrate that the normalized polarization profile $P(z/H)/P_s$ and the mean polarization $\bar{P}/P_s$ depend only on the relative film thickness $H/\xi^*$ and the ratio $\delta/\xi^*$. This result agrees with the behavior of free standing films.[6]

Consider now the variation of the film mean polarization $\bar{P}$ with temperature $T$. It results mainly from the temperature dependence of the dielectric stiffness $a_1$, which follows from the Curie-Weiss law and reads

$$a_1 = \frac{T-\theta}{2\varepsilon_0 C}, \qquad (19)$$

where $C$ and $\theta$ are the Curie-Weiss temperature and constant of a bulk crystal, and $\varepsilon_0$ is the permittivity of the vacuum. Using Eqs. (4), (15), and (19), and neglecting the temperature dependence of the misfit strain $S_m$, we find the renormalized correlation length $\xi^*$ as a function of the relative temperature $t = T/\theta^*$:

$$\xi^* = \sqrt{\frac{g_{11}}{|a_3^*|}} = \sqrt{\frac{2\varepsilon_0 C g_{11}}{\theta^* - T}} = \xi_0^* (1-t)^{-1/2}, \qquad (20)$$

$$\xi_0^* = \sqrt{\frac{2\varepsilon_0 C g_{11}}{\theta^*}}, \qquad (21)$$

where $\theta^*$ is the Curie-Weiss temperature of a thick epitaxial film grown on a compressive substrate, which is given by[11]

$$\theta^* = \theta + S_m \frac{4\varepsilon_0 C Q_{12}}{s_{11} + s_{12}}. \qquad (22)$$

It should be noted that here, as usual, the material constants $Q_{12}$, $s_{11}$, $s_{12}$, and $g_{11}$ are regarded as temperature-independent parameters. When the dielectric stiffness coefficient $a_{11}$ is also independent of temperature like in PT, we may write the following expression for the thick-film spontaneous polarization $P_s$:

$$P_s = \sqrt{\frac{\theta^* - T}{4\varepsilon_0 C a_{33}^*}} = P_s^0 (1-t)^{1/2}, \qquad (23)$$

$$P_s^0 = \sqrt{\frac{\theta^*}{4\varepsilon_0 C a_{33}^*}}. \qquad (24)$$

Substituting $P(z/\xi^*) = P_s \rho(\zeta)$ into Eq. (12) and using Eqs. (20) and (23), one can calculate numerically the relative mean polarization $\bar{P}(T)/P_s^0$ via the integral relation

$$\frac{\bar{P}}{P_s^0} = \frac{2\xi_0^*}{H} \int_0^{\frac{H\sqrt{1-t}}{2\xi_0^*}} \rho(\zeta)d\zeta. \qquad (25)$$

From Eq. (25) it follows that $\bar{P}(T)/P_s^0$ is a function of the relative temperature $T/\theta^*$, normalized film thickness $H/\xi_0^*$, and the ratio $\delta/\xi^*$. We note that the polarization profile $\rho(\zeta)$ depends on the relative film thickness $H/\xi^*(T)$ and so varies with temperature. The analysis shows that the extrapolation length $\delta$ should be proportional to the correlation length $\xi^*(T)$ in order to make the thermodynamic model self-consistent. Therefore, the ratio $\delta/\xi^*$ is assumed here to be independent of temperature. (Similar assumption was made by Scott et al.,[5] whereas Tilley and Žekš[4] regarded $\delta$ as a temperature-independent parameter.)

The universal temperature dependences of the film relative polarization $\bar{P}/P_s^0$, calculated on the basis of Eq. (25), are shown in Fig. 1 for films of various thicknesses $H/\xi_0^*$ grown at zero misfit strain ($S_m = 0$).

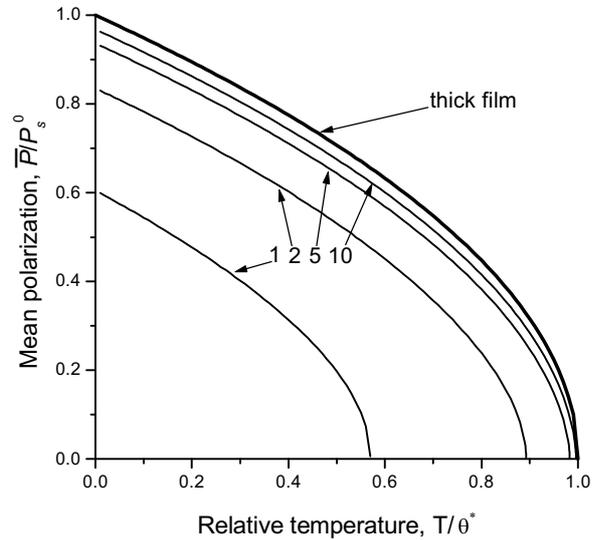

**Fig. 1.** Universal temperature dependences of the relative mean polarization $\bar{P}/P_s^0$ in epitaxial thin films experiencing a second-order ferroelectric phase transition. Films with various relative thicknesses $H/\xi_0^*$ indicated in the figure are assumed to be grown at $S_m = 0$. The curves here and below were computed at a value of $\delta/\xi^* = 1.41$ taken from Ref. 6. The upper curve shows the polarization calculated in the thick-film approximation, which neglects the surface effect.

It can be seen that the surface effect hinders the development of spontaneous polarization during the film cooling. This intrinsic size effect manifests itself in a decrease of the ferroelectric transition temperature $T_c$, as well as in a reduction of the mean polarization $\bar{P}(T)$ relative to the thick-film polarization $P_s(T)$. However, $\bar{P}(T)$ cannot be strongly suppressed in a wide temperature range, i.e. well below $T_c(H)$ the mean polarization has the same order of magnitude as $P_s(T)$. The decrease of the transition temperature relative to the thick-film value of $T_c = \theta^*$ becomes significant only in films with $H<5\xi_0^*$. Remarkably, at $H>10\xi_0^*$ the transition temperature becomes indistinguishable from $\theta^*$ (see Fig. 1), which supports the validity of calculations performed earlier in the thick-film approximation.[11]

When the mean polarization $\bar{P}(T)$ is normalized by the temperature-dependent thick-film polarization $P_s(T)$, it also exhibits a universal dependence on the relative film thickness $H/\xi^*(T)$. This dependence, calculated in the $P^4$- approximation, is presented in the Fig. 2. It can be seen that, irrespective of temperature, strong suppression of the film polarization takes place only in ultrathin films with a thickness $H$ comparable to the correlation length $\xi^*(T)$.

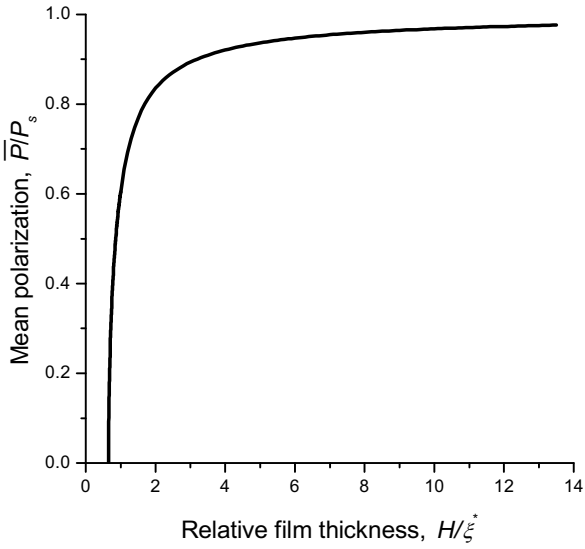

**Fig. 2.** Universal dependence of the normalized mean polarization $\bar{P}(T)/P_s(T)$ on the relative film thickness $H/\xi^*(T)$. The curve was calculated in the $P^4$- approximation for epitaxial films grown at $S_m = 0$.

For PT and BT films, temperature dependences of the mean polarization $\bar{P}(T)$ were also calculated with the aid of Eqs. (10)-(12), which take into account the sixth-order term in the free-energy expansion. Numerical values of the involved material parameters were taken from Ref. 18. (The coefficient $g_{11}$, which defines the gradient term in the energy expansion, does not affect the dependence $\bar{P}(T)$ plotted at a given normalized thickness $H/\xi_0^*$.) In the case of PT films, the misfit strain in the epitaxial system was assumed to be temperature-independent for simplicity ($S_m = 0$). For BT films, it was given a linear temperature dependence[23] corresponding to a representative example of BT films grown on MgO [24] in order to avoid the transformation of the tetragonal phase into a monoclinic one ($P_1 = P_2 \neq 0$, $P_3 \neq 0$) during the cooling.[11,18]

Figure 3 shows the mean polarization $\bar{P}(T)$ calculated for PT films as a function of temperature. For comparison, the dependences $\bar{P}(T)$ obtained in the $P^4$-approximation are also plotted in Fig. 3(a). It can be seen that in the vicinity of the transition temperature $T_c(H/\xi_0^*)$ the $P^6$-term indeed may be neglected. However, well below $T_c$ the $P^4$-approximation overestimates the surface effect considerably. With decreasing temperature, the actual polarization $\bar{P}(T)$ first increases more steeply and then varies more slowly than predicted by the $P^4$-model.

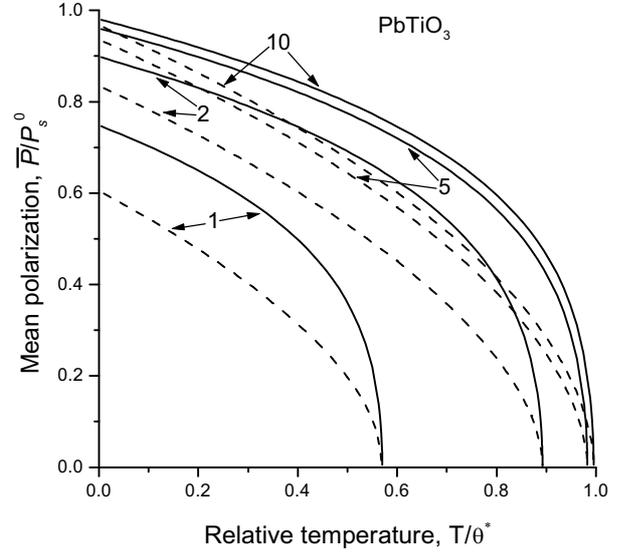

(a)

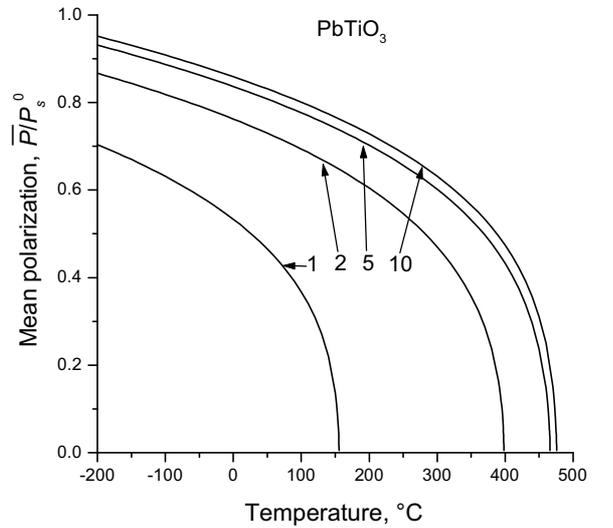

(b)

**Fig. 3.** Mean polarization $\bar{P}$ of epitaxial PbTiO$_3$ films grown at $S_m = 0$ as a function of the relative (a) and actual (b) temperature. Solid curves were calculated with the account of the $P^6$-term in the energy expansion, whereas dotted lines show results obtained in the $P^4$-approximation. The curves $\bar{P}(T)$ were normalized by the thick-film polarization $P_s^0(T = 0)$ calculated in the corresponding approximation. Numbers indicate the normalized film thickness $H/\xi_0^*$ for respective curves.

As follows from Fig. 4, the BT films display a similar temperature dependence $\overline{P}(T)$. Remarkably, the negative surface effect cannot strongly suppress the film polarization $\overline{P}(T)$ in a wide temperature range, though it may reduce the transition temperature $T_c$ significantly. In view of this result, the observation of a very low remanent polarization (less than the bulk value by an order of magnitude) in ultrathin BT films ($H \approx$ 12 nm) grown on SrTiO$_3$ by the molecular beam epitaxy[26] cannot be attributed to the intrinsic size effect.

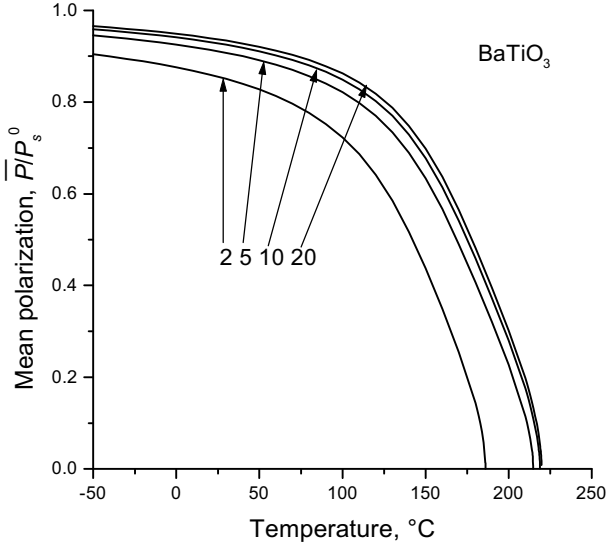

**Fig. 4.** Temperature dependences of the normalized mean polarization in BaTiO$_3$ films grown on MgO at $T_g = 780°$ C. Films are assumed to be fully relaxed ($S_m = 0$) at 780° C. Mean thermal expansion coefficients are taken to be $10.4 \times 10^{-6}$ K$^{-1}$ (BT) and $14.8 \times 10^{-6}$ K$^{-1}$ (MgO).[25] The mean polarization $\overline{P}$ is normalized by a polarization $P_s^0(T = 0)$ calculated in the thick-film approximation at a value of the misfit strain $S_m = -4.6 \times 10^{-3}$, which would appear in this film/substrate system at $T = 0$. Numerical values of the film thickness $H/\xi_0$ normalized by the bulk correlation length $\xi_0 = \xi(T = 0)$ are indicated in the figure.

## IV. FILM POLARIZATION: MISFIT-STRAIN DEPENDENCE

We proceed now to the description of the misfit-strain effect on ferroelectricity in ultrathin films. As shown earlier,[11,18] the transition temperature $T_c$ of a thick epitaxial film is close to the Curie temperature $T_c^b$ of a bulk crystal only in the vicinity of $S_m = 0$. At large negative misfit strains, which are provided by strongly compressive substrates, $T_c(S_m)$ may rise markedly with respect to $T_c^b$. In ultrathin films, an additional negative surface effect appears, tending to reduce $T_c$. Therefore, the shift of the transition point relative to $T_c^b$ is governed in ultrathin epitaxial films by the competition of the strain and surface effects on $T_c$.

Figure 5 shows the transition temperature $T_c$ of PT films calculated as a function of the misfit strain $S_m$ at different values of the relative film thickness $H/\xi_0$. [Here $H$ is normalized by the bulk correlation length $\xi_0 = \xi(T = 0)$, which coincides with $\xi_0^*(S_m = 0)$]. It can be seen that $T_c$ varies linearly with $S_m$, irrespective of the film thickness. The slope of this linear dependence decreases only slightly in ultrathin films in comparison with thick ones. Remarkably, even in PT films with a small thickness of $H = 5\xi_0$, the transition temperature appears to be higher than $T_c^b$ in a wide misfit-strain range of $S_m < -1 \times 10^{-3}$ (see Fig. 5). This is due to the fact that the surface effect reduces $T_c$ by less than 20° C in these films. At the same time, the strain-induced increase of $T_c$ at $S_m = -5 \times 10^{-3}$, for example, is about 100 °C. Owing to the dominant role of the misfit-strain effect, significant suppression of the transition temperature is expected to occur only in ultrathin films (thickness about $2\xi_0$ or less) at small misfit strains.

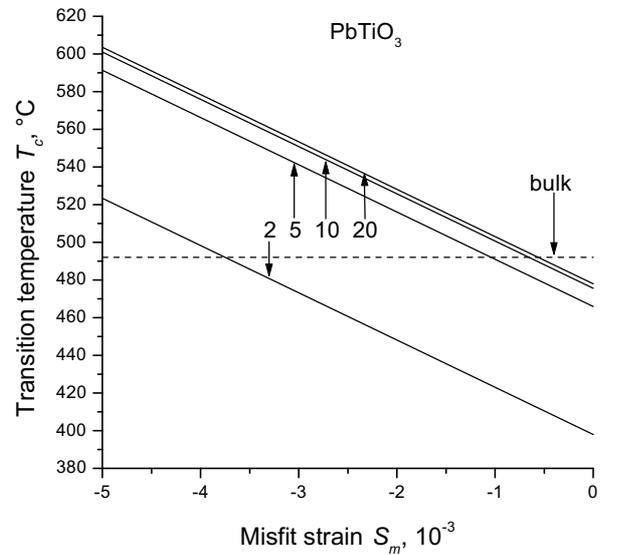

**Fig. 5.** Ferroelectric transition temperature $T_c$ calculated as a function of the misfit strain $S_m$ for PbTiO$_3$ films of several different thicknesses. Numbers indicate the normalized film thickness $H/\xi_0$ for respective curves. Dashed line shows the transition temperature of a bulk crystal.

To compare our theoretical predictions with experimental data, we must determine the magnitude of the ferroelectric correlation length $\xi_0^*$. According to Eq. (21), $\xi_0^*$ depends on the gradient coefficient $g_{11}$ involved in the free-energy expansion (3). This coefficient may be evaluated from the observed widths of domain walls in ferroelectrics using the advanced theory of domain boundaries developed by Cao and Cross.[16] For 90° domain walls, which are thick in comparison with 180° ones, the wall width at room temperature was found to be about 1 nm in PT and 5 nm in BT.[27,28] Using these values and Eq. (5.15a) derived in Ref. 16 for the half-width of a 90° domain wall, we calculated the difference between $g_{11}$ and the other gradient coefficient $g_{12}$ to be $g_{11} - g_{12} \approx 3.5 \times 10^{-10}$ J·m$^3$/C$^2$ in PT and $g_{11} - g_{12} \approx 2.7 \times 10^{-9}$ J·m$^3$/C$^2$ in BT.[29] Assuming that $g_{12}$ is negligible in comparison with $g_{11}$ and using Eq. (21), we evaluated the correlation length $\xi_0^*(S_m = 0) = \xi(T = 0)$ to be about 1 nm in PT and 5 nm in BT. Substituting Eq. (22) into Eq. (21), we also calculated the misfit-strain dependence of the renormalized correlation length $\xi_0^*$. Numerical results obtained for PT and BT films are shown in Fig. 6. It can be seen that the two-dimensional compression reduces the correlation length and, therefore, partly suppresses the surface effect.

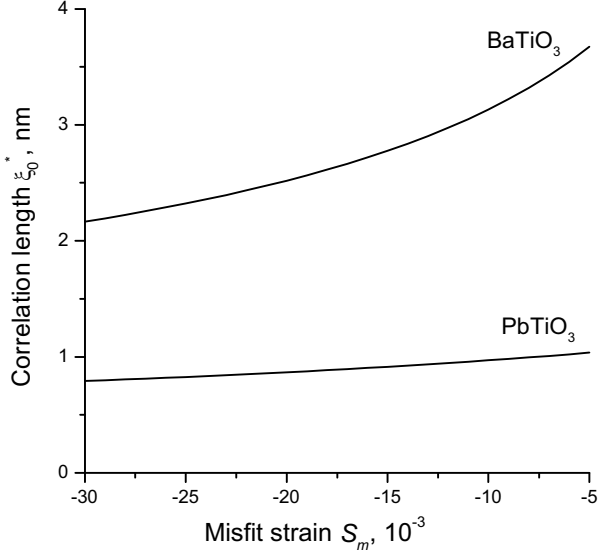

**Fig. 6.** Misfit-strain dependence of the renormalized ferroelectric correlation length $\xi_0^*$ in epitaxial PbTiO$_3$ and BaTiO$_3$ films.

Since in PT $\xi_0^*(S_m = 0) = \xi_0$ is about 1 nm, from Fig. 5 it follows that even in ultrathin epitaxial PT films with $H \approx 2$ nm the transition temperature is expected to be much higher than the room temperature. This theoretical result agrees with the observation of a stable out-of-plane polarization in 4-nm-thick Pb(Zr$_{0.2}$Ti$_{0.8}$)O$_3$ films grown on SrTiO$_3$, which was reported in Ref. 12. (Though our mean-field phenomenological theory is probably not valid for ultrathin films with thicknesses about 1 nm and below, we believe that films containing 10 or more monolayers can be successfully described in the continuum approximation.)

Finally, we shall consider the misfit-strain dependence of the mean polarization $\overline{P}$ in ultrathin films at room temperature. Figure 7 shows this dependence calculated for BT films of different relative thicknesses $H/\xi_0$. It can be seen that the substrate-induced lattice compression leads to a significant increase of the mean polarization $\overline{P}$ not only in thick films ($H \gg \xi_0$), but also in ultrathin epitaxial layers ($H \sim \xi_0$). At large negative misfit strains $S_m < -10 \times 10^{-3}$, the strain effect completely overrides the polarization suppression caused by the surface effect. As a result, in a wide misfit-strain range the film mean polarization becomes larger than the spontaneous polarization $P_s = 0.26$ C/m$^2$ of bulk BT (see Fig. 7).

Our theoretical predictions may be compared with the experimental data of Yanase et al. on the polarization properties of epitaxial BT films grown by radio-frequency magnetron sputtering on SrRuO$_3$-covered SrTiO$_3$ crystals.[13] Clear ferroelectric hysteresis loops were observed by these authors even in ultrathin BT films with $H = 12$ nm. The remanent polarization $P_r$ was found to be larger than the bulk polarization in the whole range of film thicknesses obtained ($H = 12$–79 nm). The maximum value of $P_r \approx 0.385$ C/m$^2$ was observed at $H = 40$ nm. According to the performed x-ray measurements,[13] the film in-plane lattice parameter $a$ at $H = 12$ nm is equal to the lattice constant $b$ of the SrTiO$_3$ substrate, whereas at $H = 59$ nm it deviates from $b$, indicating a decrease of the compressive strain in the film. Therefore, gradual reduction of $P_r$ with increasing film thickness at $H > 40$ nm can be attributed to the strain relaxation caused by the generation of misfit dislocations at the growth temperature. The decrease of $P_r$ at $H < 40$ nm may be regarded as a manifestation of the negative surface effect (see Fig. 7).

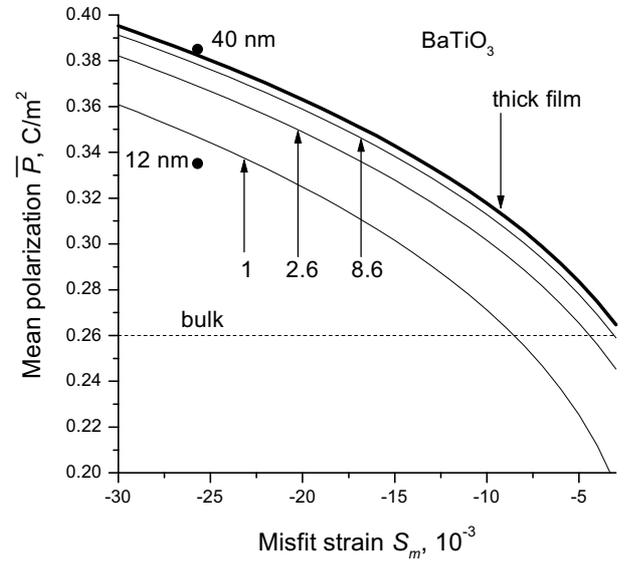

**Fig. 7.** Misfit-strain dependences of the mean polarization in BaTiO$_3$ films of different thicknesses at room temperature. Numbers indicate the normalized film thickness $H/\xi_0$ for respective curves. Circles show the remanent polarizations, which were observed in BT films epitaxially grown on SrRuO$_3$ covered SrTiO$_3$ crystals (Ref. 13). Dashed line shows the spontaneous polarization of a bulk crystal.

Indeed, the minimum observed remanent polarization of $P_r(H = 12$ nm$) = 0.335$ C/m$^2$ is considerably lower than the thick-film polarization $\overline{P} = 0.383$ C/m$^2$ calculated at the misfit strain of $S_m = -26 \times 10^{-3}$, which corresponds to ultrathin BT films grown on SrTiO$_3$.[30] At the same time, the polarization $P_r \approx 0.385$ C/m$^2$ of the 40-nm-thick BT film is in good agreement with the theoretical value of $\overline{P} = 0.379$ C/m$^2$ obtained for the respective normalized thickness ($H/\xi_0 = 8.6$ at $\xi_0 = 4.65$ nm). Thus, our thermodynamic theory gives a reasonable explanation for the thickness dependence of the remanent polarization, which was observed in ultrathin BT films grown on SrRuO$_3$-covered SrTiO$_3$ crystals.

## V. CONCLUSIONS

1) A renormalization of the ferroelectric correlation length $\xi$ takes place in epitaxial films due to the straining of the film by a thick dissimilar substrate. The renormalized correlation length $\xi^*$ is a function of the misfit strain $S_m$ in the heterostructure. The magnitude of $\xi^*$ in the films deposited on compressive substrates ($S_m < 0$) is reduced relative to the bulk value. Accordingly, the intrinsic surface effect on ferroelectricity becomes weaker in epitaxial films grown on strongly compressive substrates.

2) In ultrathin epitaxial films, the ferroelectric transition temperature $T_c$ may be shifted from the bulk one both to higher and lower temperatures. The shift of the transition point is governed by the competition of the misfit-strain and surface effects on $T_c$. Though the negative surface effect may decrease $T_c$ considerably, it cannot strongly suppress the film mean polarization $\overline{P}(T)$ in a wide temperature range.

3) Well below the transition temperature $T_c$, the film mean polarization $\overline{P}$ may be larger than the bulk polarization even in an ultrathin epitaxial layer. This increase is caused by the in-plane straining of the film lattice by a compressive substrate, which overrides the negative surface effect.

**ACKNOWLEDGMENT**

The research described in this publication was made possible in part by Grant No. I/75965 from the Volkswagen-Stiftung, Germany. Part of the work was supported by the Strategiefonds "Piccolo" of the Helmholtz-Gesellschaft (HGF).